\documentclass[twoside,english]{jfm}
\usepackage[T1]{fontenc}
\usepackage[latin9]{inputenc}
\usepackage[active]{srcltx}
\usepackage{verbatim}
\usepackage{ifsym}
\usepackage{amsmath}
\usepackage{amssymb}
\usepackage{wasysym}
\usepackage{graphicx}
\usepackage{esint}

\makeatletter
\numberwithin{equation}{section}
\numberwithin{figure}{section}

\usepackage{graphicx}
\usepackage{natbib}

\ifCUPmtlplainloaded \else
  \checkfont{eurm10}
  \iffontfound
    \IfFileExists{upmath.sty}
      {\typeout{^^JFound AMS Euler Roman fonts on the system,
                   using the 'upmath' package.^^J}%
       \usepackage{upmath}}
      {\typeout{^^JFound AMS Euler Roman fonts on the system, but you
                   dont seem to have the}%
       \typeout{'upmath' package installed. JFM.cls can take advantage
                 of these fonts,^^Jif you use 'upmath' package.^^J}%
      }
  \else
  \fi
\fi

\ifCUPmtlplainloaded \else
  \checkfont{msam10}
  \iffontfound
    \IfFileExists{amssymb.sty}
      {\typeout{^^JFound AMS Symbol fonts on the system, using the
                'amssymb' package.^^J}%
       \usepackage{amssymb}%
         \let\leq=\leqslant
         \let\geq=\geqslant
      }{}
  \fi
\fi

\ifCUPmtlplainloaded \else
  \IfFileExists{amsbsy.sty}
    {\typeout{^^JFound the 'amsbsy' package on the system, using it.^^J}%
     \usepackage{amsbsy}}
    {\providecommand\boldsymbol[1]{\mbox{\boldmath $##1$}}}
\fi

\newsavebox{\astrutbox}
\sbox{\astrutbox}{\rule[-5pt]{0pt}{20pt}}

\title[Properties of the compact Zakharov equation]{On certain properties of the compact Zakharov equation}

\author[F. Fedele]%
{Francesco Fedele%
  \thanks{Email address for correspondence: E-mail address: fedele@gatech.edu}}

\affiliation{Address: School of Civil and Environmental Engineering,\\
School of Electrical and Computer Engineering,\\
Georgia Institute of Technology, Atlanta, Georgia, USA}

\pubyear{2013}
\volume{xxx}
\pagerange{xxx--xxx}
\date{?; revised ?; accepted ?. - To be entered by editorial office}

\makeatother

\usepackage{babel}
\begin{document}
\maketitle\global\long\def\S{\mathcal{S}}
\global\long\def\eps{\varepsilon}
\global\long\def\H{\mathcal{H}}
\global\long\def\L{\mathcal{L}}
\global\long\def\M{\mathcal{M}}
\global\long\def\K{\mathbf{K}}
\global\long\def\Hilb{\mathbf{H}}
\global\long\def\R{\mathbb{R}}
\global\long\def\ud{\mathrm{d}}

\begin{abstract}
Long-time evolution of a weakly perturbed wavetrain near the modulational
instability threshold is examined within the framework of the compact
Zakharov equation for unidirectional deep-water waves (\cite{Dyachenko2011}).
Multiple-scale solutions reveal that a perturbation to a slightly
unstable uniform wavetrain of steepness $\mu$ slowly evolves according
to a nonlinear Schrodinger equation. In particular, for small carrier
wave steepness $\mu<\mu_{1}\approx0.27$ the perturbation dynamics
is of focusing type and the long-time behavior is characterized by
the Fermi-Pasta-Ulam recurrence, the signature of breather interactions.
However, the amplitude of breathers and their likelihood of occurrence
tend to diminish as $\mu$ increases while the Benjamin-Feir index
decreases and becomes nil at $\mu_{1}$. This indicates that homoclinic
orbits persist only for small values of wave steepness $\mu\ll\mu_{1}$,
in agreement with recent experimental and numerical observations of
breathers.

When the compact Zakharov equation is beyond its nominal range of
validity, i.e. for $\mu>\mu_{1}$, predictions seem to foreshadow
a dynamical trend to wave breaking. In particular, the perturbation
dynamics becomes of defocusing type, and nonlinearities tend to stabilize
a linearly unstable wavetrain as the Fermi-Pasta-Ulam recurrence is
suppressed. At $\mu=\mu_{c}\approx0.577$, subharmonic perturbations
restabilize and superharmonic instability appears, possibly indicating
that wave dynamical behavior changes at large steepness, in qualitative
agreement with the numerical simulations of \cite{Longuet-HigginsCokelet1978}
for steep waves. Indeed, for $\mu>\mu_{c}$ a multiple-scale perturbation
analysis reveals that a weak narrowband perturbation to a uniform
wavetrain evolves in accord with a modified Korteweg-de Vries/Camassa-Holm
type equation, again implying a possible mechanism conducive to wave
breaking. \end{abstract}
\begin{keywords}
Breathers, Hamiltonian, modulational instability, multiple-scale perturbation,
nonlinear waves, recurrence, wave breaking.
\end{keywords}

\section{Introduction}

Unidirectional weakly nonlinear narrowband wavetrains evolve in deep
water according to the nonlinear Schrodinger (NLS) equation, which
is integrable. The associated Lax-pairs were discovered by \cite{Zakharov1972},
who unveiled the dynamics of solitons via the Inverse Scattering Transform
(IST) (see e.g. \cite{Ablowitz1981}). Another important asymptotic
model of the Euler equations for the free-surface of an ideal flow
is the Zakharov (Z) integro-differential equation, which is not integrable
(\cite{Zakharov1999} and \cite{Dyachenko2013}). The Z equation is
derived by means of a third order expansion of the Hamiltonian in
wave steepness, where fast non-resonant interactions are eliminated
via a canonical transformation that preserves the Hamiltonian structure
(\cite{Krasitskii1994}). The equation is valid for weakly nonlinear
four-wave interactions, but it has no constraints on the spectral
bandwidth. For unidirectional waves with narrowband spectra it reduces
to the NLS or the higher order \cite{Dysthe1979} equation.

It is well known that a finite-amplitude uniform wavetrain is unstable
to infinitesimal subharmonic perturbations, the so-called modulational
instability (MI) or Benjamin-Feir instability (\cite{Benjamin1967a,Benjamin1967}).
Whereas the MI growth rate implied by the NLS model tends to overestimate
experimental data, growth rates predicted from the Z equation are
lower and comparable to the values observed in experiments (\cite{Crawfordetal1981,Janssen1983}).
Further, \cite{JanssenPoF1981} showed within the NLS framework that
in the absence of viscous dissipation, a linearly unstable wavetrain
does not evolve to a steady state, but the long-time behavior is characterized
by successive modulation and demodulation cycles, viz. the Fermi-Pasta-Ulam
(FPU) recurrence (\cite{FermiPastaUlam}). This is the signature of
breathers, homoclinic orbits to an unstable uniform wavetrain (\cite{Peregrine1983}
and \cite{Osborne2010}, see also \cite{Henderson1999341} and \cite{Tanaka1990559}). 

NLS breathers have been the subject of numerous studies, in particular,
to explain rogue wave formation (\cite{Dysthe1999,Osborne2000,Peregrine1983,Kharif2009a,Kharif2003,Janssen2003,GramstadTrulsen2007,DystheKrogstad2008,Clamond2006}).
Recently, \cite{Chabchoub2011} and \cite{Chabchoubc2012} provided
laboratory observations of higher-order breathers at sufficiently
small values of wave steepness ($\sim0.01-0.09$), confirmed by numerical
simulations (\cite{Slunyaev2013PRE}). 

The experimental and numerical results describing the nature of breathers
as briefly reviewed in the preceding provide the principal motivation
for this study. In particular, we aim to investigate further the modulational
properties of the Z equation for unidirectional deep-water waves.
This should provide new insight into the occurrence of breathers as
observed in the experimental studies aforementioned. To achieve this
objective, we shall study the weakly nonlinear space-time evolution
of a unstable uniform wavetrain of the compact Z equation, hereafter
referred to as cDZ. The compact form follows from a canonical transformation
of the Z equation and eliminates trivial resonant quartet interactions
(\cite{Dyachenko2011}). As a result, the Z model reduces to a generalized
derivative NLS type equation (\cite{FedeleDutykhJFM2012}). 

The long-time behavior near the MI threshold is determined by means
of multiple-scale perturbation techniques (see e.g. \cite{Yang2010}).
Although the cDZ equation is strictly valid for weakly nonlinear four-wave
interactions, it captures new features that indicate finite-time blowup
or wave breaking, not modeled by the one-dimensional (1D) NLS nor
the higher-order \cite{Dysthe1979} equation. Therefore we shall also
explore its properties for relatively large steepness values beyond
the range of validity since the resulting predictions may serve to
indicate the behavior of waves as they steepen and approach breaking.

The remainder of the paper is organized as follows. In section 2,
the cDZ equation is introduced, and then the associated equations
in terms of local wave amplitude and phase are derived. The linear
instability of a uniform wavetrain is presented in section 3 and followed
by a multiple-scale perturbation analysis to study the long-time dynamics
of a weakly perturbed wavetrain in section 4. This is followed by
a discussion of the theoretical results and concluding remarks.

\section{Compact Zakharov equation}

Following \cite{FedeleDutykhJFM2012}, we introduce a reference frame
moving at the group velocity $c_{g}=\omega_{0}/(2k_{0})$ in deep
water and the dimensionless scales $X=k_{0}(x-c_{g}t)$ and $T=\omega_{0}t$,
with $k_{0}=\omega_{0}^{2}/g$ and $\omega_{0}$ as the wavenumber
and frequency of the carrier wave $e^{i(k_{0}x-\omega_{0}t)}$. The
leading order wave surface $\eta$ is given by 
\begin{equation}
k_{0}\eta(X,T)=B(X,T)e^{i(k_{0}x-\omega_{0}t)}+\mathrm{c.c.,}\label{eq:etaEnv}
\end{equation}
and non-dimensional envelope $B$ follows from

\begin{equation}
i\partial_{T}B=\frac{\delta\H}{\delta B^{*}},\label{eq:env}
\end{equation}
where $\delta$ denotes variational differentiation, 
\begin{equation}
\H=\int_{\R}\Bigl[B^{\ast}\Omega B+\frac{i}{4}|\S B|^{2}[B(\S B)^{\ast}-B^{\ast}\S B]-\frac{1}{2}|\S B|^{2}\Hilb(\partial_{X}|B|^{2})\Bigr]\,\ud X\label{eq:H}
\end{equation}
is the Hamiltonian and

\[
\S=\partial_{X}+i,\qquad\Omega=\frac{1}{8}\partial_{XX},
\]
with $\Hilb(g)$ being the Hilbert transform of $g(X)$. The preceding
do not include third- and higher-order corrections to dispersion to
simplify the analysis of the cDZ. Further, higher-order non-resonant
corrections to (\ref{eq:env}) hidden within the full canonical transformation
of \cite{Dyachenko2011} are not accounted for, and the carrier wave
steepness is defined as 
\begin{equation}
\mu=k_{0}a_{0}=2\left|B\right|,\label{mu}
\end{equation}
where $a_{0}=2\left|B\right|/k_{0}$ is the amplitude of $\eta$.
Also note that (\ref{eq:env}) is valid if all the Fourier components
comprising the spectrum of $\eta$ travel in the same direction (\cite{Dyachenko2011}).
This condition is satisfied if $k_{0}\gg1$ or the spectrum of $B$
has negligible energy for wavenumbers $k<-k_{0}$, viz. the spectral
bandwidth $\Delta k/k_{0}$ is less than unity. Otherwise, a projection
operator $P^{+}$ would have to be applied to the nonlinear term of
(\ref{eq:env}) to nullify Fourier modes with wavenumbers $k<-k_{0}$.
We assume that the conditions for excluding $P^{+}$are satisfied
in the present analysis.

The uniform wavetrain solution of the cDZ equation is 
\begin{equation}
B_{0}(T)=\sqrt{E_{0}}e^{-iE_{0}T},\label{eq:B0}
\end{equation}
where $E_{0}$ is the squared amplitude of the wavetrain. 

The stability of $B_{0}$ to infinitesimal perturbations and its weakly
nonlinear evolution over the long-time scale can be studied by considering
the local form of the cDZ equation, ignoring the effects of wave-induced
currents. This does not affect the eventual conclusions of the present
analysis. Under this setting, define 
\begin{equation}
B=\sqrt{E(X,T)}e^{i\phi(X,T)-iE_{0}T},\label{ansatz}
\end{equation}
with $E$ as the squared envelope amplitude and $\phi$ the associated
phase. By neglecting non-local terms in (\ref{eq:H}), the Lagrangian
$\L$ associated with (\ref{eq:env}), namely

\begin{equation}
\L=\frac{i}{2}\left(B^{*}\partial_{T}B-\partial B_{T}B^{*}\right)-\H\label{eq:L}
\end{equation}
reduces to
\begin{equation}
\L=-E\left(E_{0}+\phi_{T}\right)-\frac{4E^{2}\left(-\phi_{X}^{2}+4E\left(1+\phi_{X}\right)^{3}\right)+E_{X}^{2}\left(-1+4E\left(1+\phi_{X}\right)\right)}{32E},
\end{equation}
where subscripts denote partial derivatives with respect to $T$ or
$X$. Minimizing the action via variational differentiation of $\L$
yields the dynamical equations for $E$ and $\phi$ as

\begin{equation}
\begin{cases}
\begin{array}{c}
\partial_{T}\phi+\omega=0\\
\\
\partial_{T}E+\partial_{X}\left(VE\right)=0
\end{array}\end{cases},\qquad\label{E_phi_eq}
\end{equation}
where, in the reference frame $X$ moving at the group speed $c_{g}$,
the local frequency of the wavetrain is given by 
\begin{equation}
\begin{array}{c}
\omega=\frac{\partial\H}{\partial E}-\partial_{X}\left(\frac{\partial\H}{\partial E_{X}}\right)=\frac{E_{XX}}{16E}\left[1-4E(1+\phi_{X})\right]+\\
\\
-\frac{1}{32}\left(\frac{E_{X}}{E}\right)^{2}-\frac{\phi_{X}^{2}}{8}+E(1+\phi_{X})^{3}-E_{0}-\frac{1}{4}E_{X}\phi_{X},
\end{array}\label{omega}
\end{equation}
and the energy flux velocity by

\begin{equation}
V=\frac{1}{E}\frac{\partial\H}{\partial\phi_{X}}=-\frac{\phi_{X}}{4}\left(1-12E\right)+\frac{3}{2}E+\frac{1}{8}\left[\frac{E_{X}^{2}}{E}+12E\phi_{X}^{2}\right].\label{c}
\end{equation}
Note that if the cubic terms are neglected, (\ref{E_phi_eq}) reduces
to the NLS model (\cite{JanssenPoF1981} and \cite{Chu1970}).

Next, we can exploit the conservative nature of the system described
by (\ref{E_phi_eq}), formulated in terms of $E$ and the local wavenumber
$K=\phi_{X}$. In particular, the differentiation of the first equation
in (\ref{E_phi_eq}) leads to

\begin{equation}
\begin{cases}
\begin{array}{c}
\partial_{T}K+\partial_{X}\omega=0\\
\\
\partial_{T}E+\partial_{X}\left(VE\right)=0
\end{array}\end{cases}.\label{K_Eeq}
\end{equation}
Here, $E$ and $K$ can be interpreted as the 'density' and 'momentum'
of a gas with 'pressure' $\omega$. Moreover, their space averages
are invariants of motion.

\section{Linear stability of a uniform wavetrain}

In accord with the ansatz (\ref{ansatz}) and (\ref{E_phi_eq}), the
uniform wavetrain solution (\ref{eq:B0}) is given in terms of $E$
and $\phi$ as 

\begin{equation}
\mathbf{v}_{0}=\left[\begin{array}{c}
E_{0}\\
\\
0
\end{array}\right].\label{v0}
\end{equation}
To proceed with the linear stability analysis of $\mathbf{v}{}_{0}$,
we perturb it as 
\begin{equation}
\mathbf{v}=\mathbf{v}{}_{0}+\epsilon\mathbf{v}_{1},\label{exp1-1}
\end{equation}
where 
\[
\boldsymbol{\mathbf{v}}_{1}=\left[\begin{array}{c}
E_{1}(X,T)\\
\\
\phi_{1}(X,T)
\end{array}\right],
\]
and $\epsilon$ is a small parameter. Linearizing (\ref{E_phi_eq})
yields the vector equation 

\begin{equation}
\partial_{T}\mathbf{v}_{1}+\M_{_{0}}\mathbf{v}_{1}=0\label{v1eq}
\end{equation}
where 
\begin{equation}
\mathbf{v}_{1}=\left[\begin{array}{c}
E_{1}\\
\phi_{1}
\end{array}\right],\qquad\M_{_{0}}=\left[\begin{array}{ccc}
3E_{0}\partial_{X} &  & \frac{-E_{0}\left(1-12E_{0}\right)}{4}\partial_{XX}\\
\\
1+\frac{1-4E_{0}}{16E_{0}}\partial_{XX} &  & 3E_{0}\partial_{X}
\end{array}\right].\label{M0}
\end{equation}
The harmonic solution of (\ref{v1eq}) is 
\begin{equation}
\mathbf{v}_{1}=\left[\begin{array}{c}
ae^{i\left(kX-wT\right)}+c.c.\\
\phi_{0}
\end{array}\right],\label{v1sol}
\end{equation}
with $k$ and $w$ as dimensionless wavenumber and frequency of the
perturbation, and $a$ and $\phi$ satisfy the system 
\[
\left[\begin{array}{ccc}
i\left(3E_{0}k-w\right) &  & \frac{E_{0}k^{2}}{4}-3E_{0}^{2}k^{2}\\
\\
1+\frac{k^{2}\left(-1+4E_{0}\right)}{16E_{0}} &  & i\left(3E_{0}k-w\right)
\end{array}\right]\left[\begin{array}{c}
a\\
\\
\phi_{0}
\end{array}\right]=\left[\begin{array}{c}
0\\
\\
0
\end{array}\right].
\]
Therefore, for non-trivial solutions 
\begin{equation}
w^{2}-6E_{0}kw+\frac{E_{0}k^{2}}{4}-\frac{k^{4}}{64}+E_{0}k^{2}\left(6E_{0}+\frac{k^{2}}{4}\right)-\frac{3}{4}E_{0}^{2}k^{4}=0.\label{weq}
\end{equation}
The growth rate $\gamma$ follows from the imaginary part of $w$
as 

\begin{equation}
\gamma^{2}=-\frac{\Delta_{w}}{4}=\frac{1}{64}\left(1-12E_{0}\right)k^{2}\left[16E_{0}-\left(1-4E_{0}\right)k^{2}\right],\label{gamma}
\end{equation}
where $\Delta_{w}$ is the discriminant of (\ref{weq}). From (\ref{mu}),
(\ref{gamma}) can be written in terms of $\mu=2\sqrt{E_{0}}$ as
\begin{equation}
\gamma^{2}=\frac{1}{64}\left(1-3\mu^{2}\right)k^{2}\left[4\mu^{2}-\left(1-\mu^{2}\right)k^{2}\right].\label{gammamu}
\end{equation}
Note that $\gamma$ vanishes at the critical dimensionless wavenumber
\begin{equation}
k_{c}^{2}=\frac{16E_{0}}{1-4E_{0}}=\frac{4\mu^{2}}{1-\mu^{2}},\label{kc}
\end{equation}
and the associated frequency $w_{c}=3E_{0}k_{c}$. It is noted that
near the instability threshold $k_{c}$, the cDZ equation (\ref{eq:env})
is valid if $k_{c}^{2}<1$ for allowing Fourier modes of the associated
wave surface $\eta$ with nonnegative wavenumbers only. This yields
the upper bound $\mu_{m}=0.447$ for $\mu$, nearly the same as the
well-known Stokes limiting steepness $0.448$. Thus, the above linear
analysis is strictly valid for wave steepness $\mu<\mu_{m}$, largely
within the range of validity of the cDZ. 

Perturbations with $k<k_{c}$ are unstable as an indication of subharmonic
instability. At the critical steepness $\mu_{c}=0.577$ where $E=E_{c}=1/12\approx0.08$,
the perturbation is neutral irrespective of $k$%
\footnote{The same threshold holds if non-local effects are retained in the
linear stability analysis (\cite{Dyachenko2011}).%
}. Despite the fact that this is greater than the Stokes limiting steepness
and beyond the validity of the cDZ, the predictions may indicate the
correct behavior as pointed out by \cite{Crawfordetal1981}. At $\mu=\mu_{c}$,
modulational (subharmonic) instability disappears whereas, for $\mu>\mu_{c}$,
superharmonic instability appears. Note that for the Z equation, $\mu_{c}\approx0.5$%
\footnote{The two thresholds are slightly different because the cDZ and Z models
are given in terms of different canonical variables. %
} (\cite{Crawfordetal1981}). For steep irrotational periodic waves
of the Euler equations, superharmonic disturbances are unstable at
about $\mu_{c}\sim0.41$ (\cite{Longuet-HigginspartI1978} and \cite{Longuet-HigginsCokelet1978}).%
\begin{comment}
The preceding thresholds are in qualitative agreement with the numerical
studies carried out by , who studied the evolution of normal-mode
perturbations of In particular, these are unstable to superharmonic
disturbances at about $\mu_{c}\sim0.41$ (\cite{Longuet-HigginspartI1978}).
\end{comment}
\begin{comment}
In particular, \cite{Longuet-HigginspartI1978} showed that some superharmonic
disturbances tend to become unstable when the steepness $\mu$ of
the unperturbed wave exceeds about $0.41$.
\end{comment}
{} %
\begin{comment}
However, he also found that subharmonic disturbances become neutrally
stable as $\mu$ increases beyond about $0.346$ whereas the cDZ model
predicts linear restabilization at $\mu_{c}=0.577$. 
\end{comment}
\begin{comment}
Table 1 summarizes the various $\mu$ definitions and values obtained
for the cDZ, Z and Euler models respectively. 
\end{comment}

\begin{figure}
\centering

\includegraphics[width=0.99\textwidth]{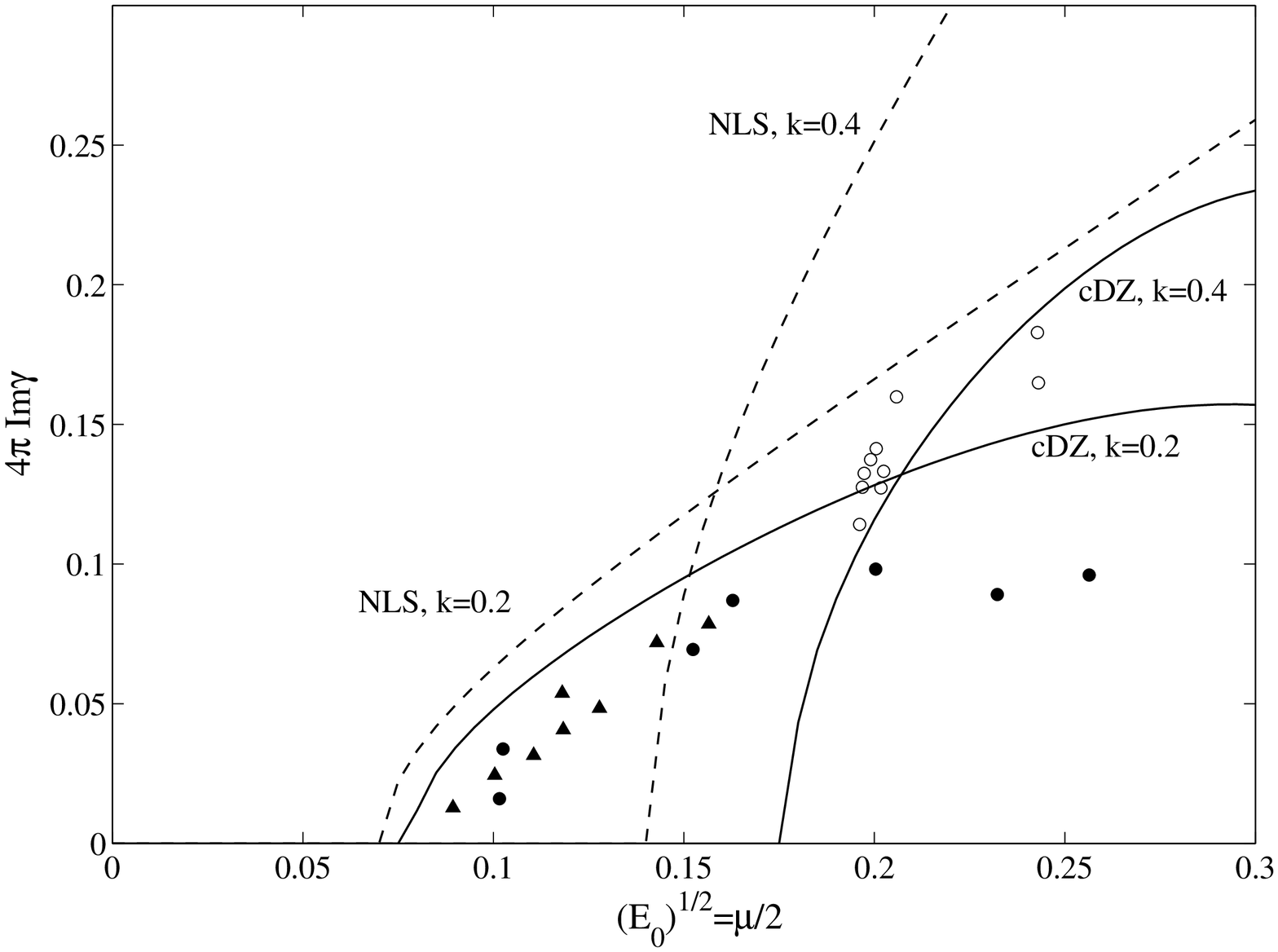} \caption{Linear sideband growth rate $4\pi\mathcal{\mathrm{Im}}(\gamma)$ of
modulated unstable wave trains as function of the amplitude $\sqrt{E_{0}}=\mu/2$
for different values of the perturbation wavenumber $k$: NLS (dashed
line) and cDZ (solid line) predictions against data digitized from
Fig. 1 in \cite{Janssen1983} (\Circle{}, $k=0.4$, \cite{lake};
\CIRCLE{}, $k=0.2$, \cite{lake}; $\text{\textifsymbol[ifgeo]{97}}$,
$k=0.2$, \cite{Benjamin1967}). }

\label{FIG1a} 
\end{figure}

In accord with cDZ, from (\ref{gammamu}) the linear growth rate $\gamma$
of a subharmonic perturbation reduces with respect to the NLS counterpart
\begin{equation}
\gamma_{NLS}^{2}=\frac{1}{64}k^{2}\left(4\mu^{2}-k^{2}\right),\label{gammamu-1}
\end{equation}
as the steepness $\mu$ of the wavetrain increases, especially for
small wavenumbers $k$'s. This is clearly seen in Fig. \ref{FIG1a},
showing the comparison of the theoretical $\gamma$ and $\gamma_{NLS}$
against laboratory data for $k=0.2$ and $0.4$ (see also \cite{Janssen1983}
and \cite{Crawfordetal1981}). This implies that MI is attenuated
as $\mu$ increases, a well established fact since \cite{LIGHTHILL01091965}
(see also \cite{Longuet-HigginspartII1978} and \cite{McLean1982}).
This indicates that the likelihood of large breathers reduces as the
carrier wave steepness increases, in agreement with the recent experimental
results on the Peregrine Breather (PB) presented by \cite{ShemerPoF2013}.
In particular, for $\mu\sim0.1$ they observed noticeable deviation
from the 1D NLS solution due to significant asymmetric spectral widening.
They reported that the ``breather does not breathe'' since no return
to the initial undisturbed wave train is observed. Moreover, the PB
amplification is slower and smaller than that predicted by the NLS
PB as an indication that MI effects attenuate as waves steepen. This
is also confirmed by recent numerical studies of the Euler equations
(\cite{SlunyaevShrira2013}).

\begin{comment}
As the perturbation evolves away from the linear regime, the diminished
linear growth of the associated amplitude induces less intense nonlinear
effects on the long-time behavior of the wave train limiting its growth,
as shown in the following section.

\begin{table}
\begin{centering}
\begin{tabular}{cccc}
\hline 
  & cDZ & Z & Euler \tabularnewline
\hline 
$\mu_{c\qquad}$ & 0.577 & 0.5 & 0.41\tabularnewline
\hline 
$\mu_{1}\qquad$ & 0.27 & - & -\tabularnewline
\hline 
$\mu_{m}\qquad$ & 0.447 & - & -\tabularnewline
\hline 
\end{tabular}
\par\end{centering}

\caption{Wave steepness thresholds relative to cDZ, Z (\cite{Crawfordetal1981})
and Euler equations (\cite{Longuet-HigginspartI1978}): superharmonic
instability occurs above $\mu_{c}$; NLS defocussing type perturbation
dynamics occurs above$\mu_{1}$ in the weakly nonlinear regime of
an unstable wavetrain; $\mu_{m}$ upper bound for the validity of
the cDZ linear analysis. }
\end{table}
\end{comment}
{} 

\section{Long-time behavior of an unstable wavetrain}

We have shown that a finite-amplitude uniform wave train is unstable
to infinitesimal perturbations of sufficiently long wavelength, viz.
$|k|<k_{c}$. Near the threshold $k_{c}$, the cDZ dynamics of a slightly
unstable wavetrain can be determined by means of multiple-scale perturbation
methods. \cite{JanssenPoF1981} has already studied the long-time
behavior of an unstable wavetrain of the NLS equation. His analysis
was restricted to the time domain and revealed that the perturbation
evolves periodically in time and exhibits FPU recurrence. 

Hereafter, in section 4.1, we extend Janssen's analysis to the cDZ
equations (\ref{E_phi_eq}). This will reveal that the perturbation
nonlinearly restabilizes as the steepness increases beyond about $\mu_{1}\approx0.27$
whereas \cite{Longuet-HigginspartI1978} found linear restabilization
of a perturbed steep periodic wave at about $0.34$. For smaller wave
steepness, the perturbation dynamics is of focusing type and dominated
by FPU recurrences. 

While the cDZ equation is only valid for broadband waves with small
steepness, the predictions based on larger values $\mu>\mu_{1}$ can
plausibly be useful in exploring their dynamics near breaking. Indeed,
the cDZ may capture new nonlinear features that are not modeled by
the 1D NLS and higher-order \cite{Dysthe1979} equations, which do
not support finite-time blowup solutions or breaking. Thus, it should
be worthwhile to explore the behavior for large values of wave steepness
in section 4.2 later on.

\subsection{Perturbation dynamics for small $\mu$}

\begin{figure}
\centering \includegraphics[clip,width=1.1\textwidth]{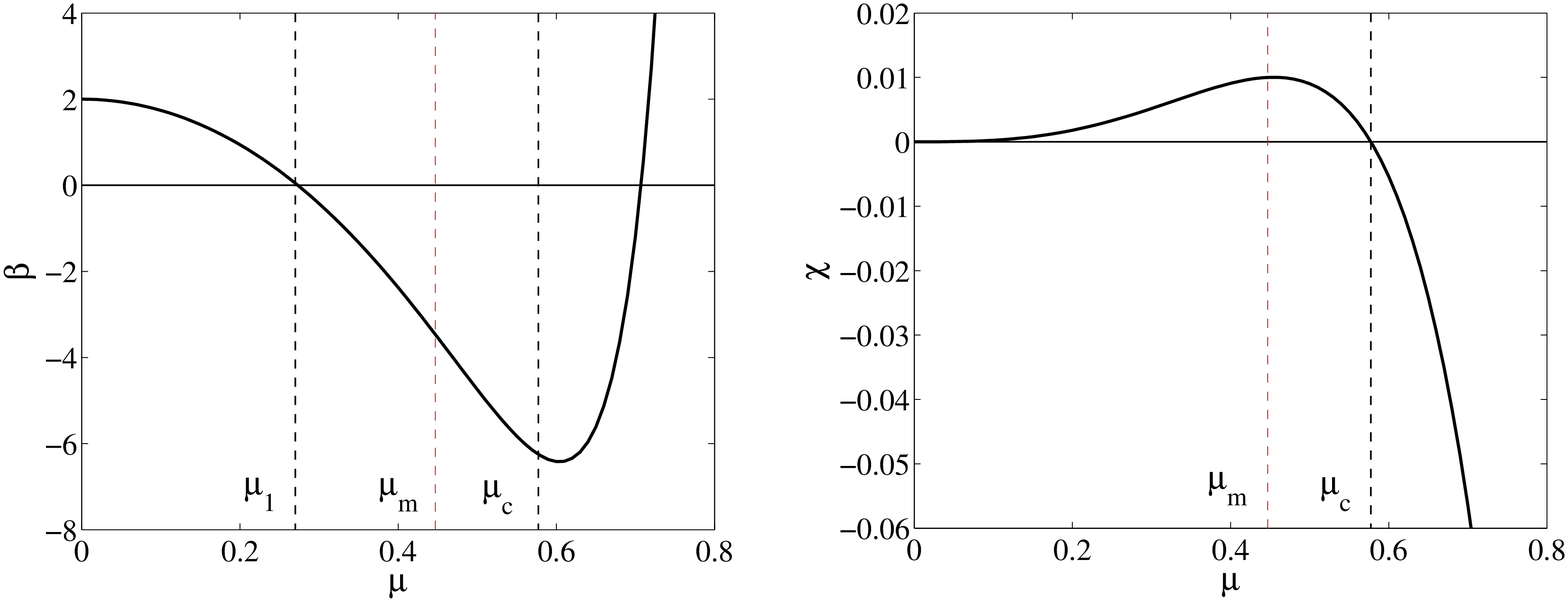}
\caption{Coefficients $\beta$ and $\chi$ of the NLS equation (\ref{NLS})
governing the perturbation to unstable wavetrain as function of wave
steepness $\mu$ ($\mu_{1}\approx0.27$, $\mu_{m}\approx0.447$ and
$\mu_{c}\approx0.577$). Note that above $\mu_{1}$ where $\beta<0$
the perturbation dynamics is of defocusing type.}

\label{FIG1} 
\end{figure}

Introduce the independent multiple scales $\xi=\epsilon^{2}X,\:\tau=\epsilon T$
and consider the ordered expansion for $E$ and $\phi$ in the small
parameter $\epsilon$ 

\begin{equation}
\mathbf{v}=\mathbf{v}{}_{0}+\epsilon\mathbf{v}_{1}(X,T,\xi,\tau)+\epsilon^{2}\mathbf{v}_{2}(X,T,\xi,\tau)+\epsilon^{3}\mathbf{v}_{3}(X,T,\xi,\tau)+...,\label{exp1}
\end{equation}
where $\mathbf{v}{}_{0}$ is given in (\ref{v0}) and 
\[
\mathbf{v}_{j}=\left[\begin{array}{c}
E_{j}(X,T,\xi,\tau)\\
\\
\phi_{j}(X,T,\xi,\tau)
\end{array}\right].
\]
The wavenumber $k$ of the perturbation is chosen just below the critical
threshold $k_{c}$ to ensure instability, viz. $k=k_{c}-\epsilon^{2}q_{e},$
and the arbitrary parameter $q_{e}>0$ is of $O(1)$. From (\ref{gamma}),
the corresponding growth rate is of $O(\epsilon)$ and given by 
\begin{equation}
\overline{\gamma}=\epsilon\sqrt{\chi q_{e}},\qquad\chi=E_{0}k_{c}\left(1-12E_{0}\right)/2,\label{grow}
\end{equation}
confirming the well-chosen slow-time scale $\tau$ (\cite{JanssenPoF1981}).
To $O(\epsilon),$ the asymptotic solution for $\mathbf{v}$ is given
by (see Appendix A) 

\[
\mathbf{v}=\left[\begin{array}{c}
E_{0}+\epsilon A(\xi,\tau)e^{i\theta}+c.c.\\
\\
\epsilon\phi_{0}(\xi,\tau)
\end{array}\right]+O(\epsilon^{2}),
\]
where $\theta=k_{c}X-w_{c}T$, the phase is described by

\[
\phi_{0}=\phi_{0}(\xi,0)-\int_{0}^{\tau}\frac{k_{c}^{2}}{16E_{0}^{2}}|A(\xi,s|^{2}ds,
\]
and the perturbation amplitude $A$ evolves in accord with the NLS
equation 

\begin{equation}
i\chi A_{\xi}=A_{\tau\tau}+\beta|A|^{2}A-\chi q_{e}A,\label{NLS}
\end{equation}
where

\begin{equation}
\beta=\frac{2\left(1-8E_{0}\right)\left(1-56E_{0}+128E_{0}^{2}\right)}{\left(1-4E_{0}\right)^{3}}=2-104E_{0}+O(E_{0}^{2}),\label{beta}
\end{equation}
Note that the linear term in (\ref{NLS}) could be removed by the
canonical transformation $A\rightarrow Ae^{iq_{e}\xi}$. Since $\mu=2\sqrt{E_{0}}$,
$\beta$ and $\chi$ can be rewritten as 

\begin{equation}
\beta=\frac{2\left(1-2\mu^{2}\right)\left(1-14\mu^{2}+8\mu^{4}\right)}{\left(1-\mu^{2}\right)^{3}}=2-26\mu^{2}+O(\mu^{3}),\label{beta&chi}
\end{equation}
and

\begin{equation}
\chi=\frac{\mu^{3}\left(1-3\mu^{2}\right)}{4\sqrt{1-\mu^{2}}}.\label{beta&chi2}
\end{equation}
The variations of $\beta$ and $\chi$ with $\mu$ are shown in Fig.
\ref{FIG1}. Correct to $O(\mu)$, $\beta=2$ and both the NLS and
Dysthe limits of the cDZ lead to the same asymptotic equation. This
limit cannot be directly compared to that by \cite{JanssenPoF1981}.
Indeed, in the latter the long-time evolution is studied near the
neutral threshold $k_{c}$ whereas, in the former the wavenumber $k$
of the perturbation is kept as a free parameter and the NLS dynamics
is studied by perturbing the coefficient of the cubic nonlinearities. 

The Benjamin-Feir Index (BFI) associated with the NLS equation (\ref{NLS})
is proportional to the coefficient $\beta$ of the cubic term. In
the focusing regime when $\beta>0$, the excess kurtosis $\lambda_{40}$
of a random perturbation is proportional to $\beta^{2}$ (\cite{Janssen2003}).
Thus, decreasing values of $\beta$ imply a smaller $\lambda_{40}$
and a reduced likelihood of large breathers. When $\beta<0$, the
NLS dynamics is of defocusing type implying suppression of the FPU
recurrence ($\lambda_{40}\leq0$) and the appearance of nonlinear
restabilization. 

To further study the perturbation dynamics as $\mu$ increases from
zero, we neglect spatial variability and simplify (\ref{NLS}) as
\[
A_{\tau\tau}+\beta|A|^{2}A-\chi q_{e}A=0.
\]
Following \cite{JanssenPoF1981}, this can be interpreted as the equation
of motion of a particle in a potential well 
\[
V(|A|)=-\frac{1}{2}\chi q_{e}|A|^{2}+\frac{1}{4}\beta|A|^{4}.
\]
Evidently, $\beta$ decreases as $\mu$ increases and vanishes at
the critical steepness $\mu_{1}\approx0.27$ (see Fig. \ref{FIG1}).
As a result, periodic solutions exist and they are given in terms
of Jacobi functions, suggesting FPU recurrence. In this case, the
NLS equation (\ref{NLS}) is of focusing type and the perturbation
$A$ evolves to a state of interacting breathers. However, their amplitude
and likelihood of occurrence are somewhat diminished because $\beta$
(and so the BFI) is a decreasing function of $\mu$, which vanishes
at $\mu_{1}$ (see Fig. \ref{FIG1}). This indicates that homoclinic
orbits persist for $\mu\ll\mu_{1}$ in agreement with the Melnikov
analysis applied to a higher-order NLS (HONLS) equation by \cite{Schober2006}.
This is also confirmed by the recent experimental observations of
higher-order breathers at sufficiently small values ($\sim0.01-0.09$)
of wave steepness (\cite{Chabchoub2011} and \cite{Chabchoubc2012}).
It is known that observing breathers in experiments where $\mu>0.1$
is difficult because of wave breaking (\cite{ShemerPoF2013} and \cite{Chabchoubc2012}).
In particular, \cite{Slunyaev2013PRE} employ sufficiently small values
$(\leq0.1)$ for $\mu$ so as to avoid wave breaking in their numerical
simulations of breathers based on the Euler equations.

\begin{comment}
{*}{*}{*}{*}{*}{*}{*}{*}{*}{*}{*}{*}{*}{*}{*}{*}{*}{*}{*}{*}{*}{*}{*}{*}
remove ???? {*}{*}{*}{*}{*}{*}{*}{*}{*}{*}{*}{*}{*}{*}{*}{*}{*}

Clearly, HONLS, which is similar to the \cite{Dysthe1979} equation,
cannot capture the reduction of the BFI as predicted by cDZ since
it does not account for nonlinear interactions induced by spectral
widening. Indeed, the cDZ dynamics predicts that the persistence of
breathers is suppressed as wave steepness increases. Indeed, as the
steepness $\mu$ increases the cDZ predicts that the linear growth
rate $\gamma\sim\sqrt{\chi}$ in (\ref{gammamu}) decreases and so
does $\beta\sim BFI$ and as result nonlinear effects induced by a
smaller perturbation's amplitude are attenuated. 
\end{comment}

Finally, where $\beta$ is positive for $\mu\geq0.707$, the perturbation
dynamics becomes of focusing type again. We have no explanation for
this, but it could simply be an artifact of the cDZ equation largely
beyond its range of validity.

\subsection{An explorative view of dynamics for $\mu>\mu_{1}$}

Despite the fact that the cDZ equation is only valid for broadband
waves with small steepness, the predictions beyond $\mu_{1}\approx0.27$
may be still indicative of the trend of wave dynamics. In particular,
from Fig. \ref{FIG1} $\beta$ is negative in the range $(\mu_{1},\mu_{c})$
and an initially unstable wavetrain restabilizes nonlinearly over
the long timescale as an indication that FPU recurrence is suppressed.
Indeed, the NLS equation (\ref{NLS}) is now of defocusing type. %
\begin{comment}
These predictions are in qualitative agreement with the numerical
results of \cite{Longuet-Higgins1978}, who found linear restabilization
of steep waves at the larger wave steepness threshold of about 0.34. 
\end{comment}

The defocusing character of the long-time perturbation evolution still
holds in the range of superharmonic instability ($\mu\geq\mu_{c}\approx0.577,$
see Fig. \ref{FIG1}). This may suggest a change in the behavior of
the cDZ dynamics as a precursor to steepening of waves and their eventual
breaking. In this regard, \cite{BridgesJFMhomoclinic} showed that
there is very simple mechanism for wave breaking near the change of
superharmonic instability. Near the change there is a homoclinic orbit
(in time) and so for some initial conditions the solution is attracted
to the slow Stokes wave whereas for other wave breaking occurs (\cite{Tanaka}
and \cite{Jillians}). 

Hereafter, we will explore the long-time evolution of a weak narrowband
perturbation to a uniform wavetrain of large amplitude $E_{0}>E_{c}$
or, equivalently large steepness $\mu>\mu_{c}$. To do so, we consider
the conservative form (\ref{K_Eeq}) of the cDZ equation and apply
the multiple-scale perturbation technique in \cite{TaniutiWeiKdV}.
This will reveal that above $E_{c}$, the weakly nonlinear dynamics
is of hyperbolic type. 

Introduce the slow multiple scales $\xi=\epsilon(X-cT),\:\tau=\epsilon^{2}T$,
with $c$ as a wave celerity to be determined and consider the ordered
expansion for the local squared envelope amplitude $E$ and wavenumber
$K$ in the small parameter $\epsilon$ as

\begin{equation}
\mathbf{w}=\mathbf{w}_{0}+\epsilon\mathbf{w}_{1}(\xi,\tau)+\epsilon^{2}\mathbf{w}_{2}(\xi,\tau)+...,\label{w}
\end{equation}
where 
\[
\mathbf{w}_{0}=\left[\begin{array}{c}
E_{0}\\
\\
0
\end{array}\right],\qquad\boldsymbol{\mathbf{w}}_{j}=\left[\begin{array}{c}
E_{j}(\xi,\tau)\\
\\
K_{j}(\xi,\tau)
\end{array}\right].
\]
For $\mu>\mu_{c}$, the leading order solution for $E$ and $K$ is
given by (see Appendix C)

\[
\mathbf{w}=\left[\begin{array}{c}
E_{0}\pm\epsilon\frac{\sqrt{-1+3\mu^{2}}}{2}F(\xi,\tau)\\
\\
\epsilon F(\xi,\tau)
\end{array}\right]+O(\epsilon^{2}),
\]
where $\mu=2\sqrt{E_{0}}$, $c=\left(3\mu\pm\sqrt{-1+3\mu^{2}}\right)/2$
and $F$ satisfies a non-dispersive Korteweg-de Vries (KdV) equation
\begin{equation}
F_{\tau}+\beta FF_{\xi}=0,\label{KdV}
\end{equation}
where

\[
\beta=3\frac{-6+8E_{0}\left(9+E_{0}\right)+\left(-1+24E_{0}\right)\sqrt{-1+12E_{0}}}{8\sqrt{2}}.
\]
Note that celerity $c$ is real for $\mu\geq\mu_{c}$ or, equivalently
$E_{0}\geq E_{c}$, rendering the dynamical equations hyperbolic%
\footnote{For $\mu<\mu_{c}$ the analysis is invalid since the system becomes
of elliptic types and $c$ is complex.%
}. This indicates that the wave dynamics is initially non-dispersive
and the slope $F_{\xi}$ blows up in finite time due to steepening.
However, (\ref{KdV}) loses its validity and the weakly nonlinear
analysis needs to be extended to higher order. If this is carried
out to $O(\epsilon^{5})$, it yields 

\begin{equation}
F_{\tau}+\beta FF_{\xi}+\epsilon\left(z_{1}F_{\xi\xi\xi}+z_{2}F^{2}F_{\xi}\right)+\epsilon^{2}z{}_{3}\left(2F_{\xi}F_{\xi\xi}+FF_{\xi\xi\xi}\right)=0,\label{CH}
\end{equation}
where
\[
z_{1}=\frac{16E_{0}\left(1-3E_{0}\right)-1}{64E_{0}},\qquad z_{2}=3\frac{-3+4E_{0}\left(9+\sqrt{-1+12E_{0}}\right)}{4},
\]
and

\[
z_{3}=\left(-1+12E_{0}\right)\frac{8E_{0}^{2}+\sqrt{-1+12E_{0}}}{128\sqrt{2}E_{0}^{2}}.
\]
This is a modified KdV/Camassa-Holm (CH) equation (\cite{Camassa1993}),
which describes the tendency of a wave perturbation to steepen and
break eventually. We point out that, recently, \cite{BridgesKdV2013}
has identified a precise mechanism for the appearance of KdV from
a NLS equation as that in (\ref{NLS}) in the defocusing regime. Thus,
KdV dynamics is potentially possible in deep water, and further studies
along this direction are desirable.

\section{Concluding remarks}

We have presented the weakly nonlinear dynamics of a perturbation
to a linearly unstable wavetrain of the cDZ equation by using multiple-scale
perturbation techniques. As wave steepness increases, the analysis
predicts that linear growth rate of an unstable perturbation and the
associated BFI both decrease, thus leading to breather suppression.
An analytical solution of the excess kurtosis $\lambda_{40}^{cDZ}$
of the wave surface described by the cDZ confirms these theoretical
results. To $O(\nu^{2})$, the excess kurtosis assumes the form (see
Appendix D) 
\[
\lambda_{40}^{cDZ}=\lambda_{40}^{NLS}\left(1-\frac{4\sqrt{3}+\pi}{8\pi}\nu^{2}\right)\approx\lambda_{40}^{NLS}\left(1-0.40\nu^{2}\right).
\]
This is smaller than the excess kurtosis 
\[
\lambda_{40}^{NLS}=\frac{\pi}{6\sqrt{3}}\frac{24\mu^{2}}{\nu^{2}}
\]
associated with the NLS equation, especially as the spectral bandwidth
widens and approaches its eventual limit $\nu$ (\cite{Janssen2003}
and \cite{Janssen2006}).

Clearly, values of $\mu$ greater than $\mu_{1}$ lie outside the
range of relative validity of the cDZ equation. Nonetheless, the predictions
indicate that above the critical threshold $\mu_{c}=0.577$, subharmonic
instability is suppressed, but linearly unstable superharmonic perturbations
arise. In comparison, the Z equation yields the same predictions for
$\mu>\mu_{c}=0.5$. The preceding thresholds are in qualitative agreement
with the stability studies of steep periodic waves for which $\mu_{c}\sim0.41$
(\cite{Longuet-HigginspartI1978} and \cite{Longuet-HigginsCokelet1978}).

Above $\mu_{c}$, the cDZ predicts a perturbation dynamics of defocusing
type and FPU recurrence is suppressed. This suggests a change in the
behavior of the cDZ dynamics above $\mu_{c}$ as a precursor to wave
steepening and eventual breaking, as suggested by \cite{Longuet-HigginspartI1978}.
Indeed, the multiple-scale analysis reveals here  that the wave dynamics
is of hyperbolic type for $\mu>\mu_{c}$. Furthermore, the long-time
evolution of a weakly nonlinear narrowband perturbation obeys a modified
KdV/CH type equation, a model that typically arises in shallow water
wave theory. This suggests that there may be physical similarities
between shallow- and deep-water waves. 

In shallow water of depth $h$ subharmonic instabilities are suppressed
at  sufficiently small depths and waves of different wavelength tend
to travel at the same speed $\sqrt{gh}$, where $g$ denotes gravitational
acceleration. That means that dispersion is suppressed and shorter
(longer) waves tend to travel faster (slower) than their linear speeds.
Furthermore, local and superharmonic instabilities are enhanced, leading
to wave breaking. 

In deep water, the cDZ equation predicts that for large steepness
$\mu>\mu_{c}\approx0.577$, subharmonic disturbances are both linearly
and nonlinearly stable, whereas superharmonic instability arises.
Moreover, dispersion is also suppressed and shorter waves tend to
travel faster than their linear speeds. Indeed, from Eq. (\ref{omega})
at the envelope maximum ($E_{X}=0,\: E_{XX}<0$) where a large crest
occurs, the local nonlinear frequency 

\begin{equation}
\,\omega_{NL}=\omega_{L}-\frac{E_{XX}}{4}(1+K)+E(1+K)^{3}\label{wnl}
\end{equation}
increases with respect to the linear counterpart

\begin{equation}
\omega_{L}=\frac{E_{XX}}{16E}-\frac{K^{2}}{8}\label{wl}
\end{equation}
as the crest steepens since both $E$ and $K$ increase. As a result,
in a reference frame moving with the group velocity of the carrier
wave, the local nonlinear phase velocity 
\begin{equation}
c_{ph,NL}=\frac{\omega_{NL}}{K}=c_{ph,L}-\frac{E_{XX}}{4K}(1+K)+E\frac{(1+K)^{3}}{K}\label{cNL}
\end{equation}
of steepening crests tends to increase with respect to their linear
counterpart
\begin{equation}
c_{ph,L}=\frac{\omega_{L}}{K}=\frac{E_{XX}}{16EK}-\frac{K}{8}.\label{cL}
\end{equation}
A similar trend is also observed for the nonlinear and linear energy
flux velocities $V_{NL}$ and $V_{L}$ beneath a crest. Indeed, from
(\ref{c})

\begin{equation}
V_{NL}=V_{L}\left(1-12E\right)+\frac{3}{2}E(1+K^{2})\label{VNL}
\end{equation}
and 

\begin{equation}
V_{L}=-\frac{K}{4}.\label{VL}
\end{equation}
As both $E$ and $K$ increase as crests steepen, $V_{NL}$ becomes
larger than $V_{L}$. 

This analysis provides insight into the physics as waves approach
conditions conducive to or near maximum recurrence and breaking within
the cDZ framework. More explicitly, consider an unsteady slowly varying
linear wave group. Wave dispersion induces a generic slowdown of the
entire wave structure as each crest in the group reaches its maximum
height. In particular, the local phase velocity varies in time and
along the group attaining its lowest value $c_{ph,L}$ at the envelope
maximum where the largest height of a crest occurs since $E_{XX}<0$
in (\ref{cL}). The crest speed also slows down, and the slowdown
is enhanced with increasing spectral bandwidth, resulting in larger
crest amplitudes. Indeed, this process causes local energy fluxes
beneath the crest to decrease since $V_{L}$ tends to diminish as
crest steepens and $K$ increases {[}see Eq.(\ref{VL}){]}. As a result,
energy flows from both the upstream and downstream regions of the
wave-group maximum, resulting in the growth of crest amplitudes. The
stronger the crest deceleration, the larger its amplitude becomes
at focus. However, as the crest grows in amplitude, the cDZ equation
predicts that nonlinearities counterbalance the linear slowdown, which
reduces by wave dispersion suppression, i.e. $c_{ph,NL}>c_{ph,L}$
as seen in Eq. (\ref{cNL}). Further, in accord with Eq. (\ref{VNL}),
the nonlinear energy flux beneath the crest tends to increase since
$V_{NL}>V_{L}$ limiting potential energy accumulation at the crest.

The preceding physical interpretation of the cDZ predictions on wave
groups suggests that dispersion suppression is the leading cause of
the observed change in behavior of the cDZ dynamics as wave steepness
increases progressively, and it may be the main physical mechanism
operative in the neighborhood of maximum recurrence or breaking. This
is supported by laboratory studies of unidirectional focusing wave
groups by \cite{Taylor1996}. At focus, they observe an increase in
the phases of high-frequency waves relative to their linear counterparts,
an indication that wave dispersion is suppressed. More recently, \cite{Banneretal2013}
carried out a multifaceted study of steep wave groups by numerical
simulations of the Euler equations, laboratory and ocean field experiments.
Their results support the preceding cDZ predictions on wave group
behavior. In particular, they found that each crest in the group decelerates,
linking the slowdown to the reduced initial speed of breaking-wave
crests (\cite{RappMelville}). 

It also appears that directional effects further enhance wave collapse,
suppressing the nonlinear focusing induced by modulational instability.
In particular, we expect the long-time evolution of a transversely
unstable wavetrain of the three-dimensional version of the cDZ equation
to obey the two-dimensional hyperbolic NLS equation 

\begin{equation}
i\chi A_{\xi}=A_{\tau\tau}-\delta A_{\zeta\zeta}+\beta|A|^{2}A-\chi q_{e}A,\label{NLS-1}
\end{equation}
where $\zeta=\epsilon Y$ is the slow scale transverse to the main
direction of propagation, and $\delta$ is a parameter that depends
on steepness and angular spreading. It is well known that (\ref{NLS-1})
can support finite-time blow-up solutions depending on the sign of
$\delta$ (\cite{Sulem1999}).

\begin{comment}
In conclusion, we believe that our theoretical results based on the
cDZ equation for unidirectional weakly nonlinear waves will be of
relevance to the recent studies on rogue waves. Moreover, the multiple-scale
analysis presented in this work provides a theoretical framework for
studying wave steepening and breaking in non-integrable asymptotic
models of the Euler equations.
\end{comment}

\section*{Acknowledgements}

FF is grateful to Profs. Michael Banner and M. Aziz Tayfun for useful
suggestions during the preparation of the manuscript. FF also thanks
Profs. Panayotis Kevrekidis, Taras Lakoba and Jianke Yang for useful
discussions on the subject of nonlinear waves and multiple-scale perturbation
methods. Financial support is gratefully acknowledged from the National
Ocean Partnership Program, through the U.S. Office of Naval Research
(Grant BAA 09-012), in partnership with Ifremer.

\section{Appendix A: Multiple-scale perturbation analysis near the neutral
threshold $k_{c}$}

Define the slow multiple-scales $\xi=\epsilon^{2}X,\:\tau=\epsilon T$,
and space and time derivatives are now given by 

\[
\frac{\partial}{\partial T}=\frac{\partial}{\partial T}+\epsilon\frac{\partial}{\partial\tau},\qquad\frac{\partial}{\partial X}=\frac{\partial}{\partial X}+\epsilon^{2}\frac{\partial}{\partial\xi}.
\]
Substitution of the ordered expansion (\ref{exp1}) into the set of
the main equations (\ref{E_phi_eq}) yields the following hierarchy
of vector equations 

\begin{equation}
O(\epsilon):\qquad\partial_{T}\mathbf{v}_{1}+\M_{_{0}}\mathbf{v}_{1}=0\label{mu1}
\end{equation}

\begin{equation}
O(\epsilon^{2}):\qquad\partial_{T}\mathbf{v}_{2}+\M_{_{0}}\mathbf{v}_{2}=-\partial_{\tau}\mathbf{v}_{1}+\mathbf{R}_{2}(\mathbf{v}_{1}),\label{mu2}
\end{equation}

\begin{equation}
O(\epsilon^{3}):\qquad\partial_{T}\mathbf{v}_{3}+\M_{0}\mathbf{v}_{3}=-\partial_{\tau}\mathbf{v}{}_{2}-\partial_{\tau\tau}\mathbf{v}_{1}-\M_{3}\mathbf{v}_{1}+\mathbf{R}_{3}(\mathbf{v}_{1},\mathbf{v}_{2})\label{mu3}
\end{equation}
where the linear differential matrix operator 
\[
\M_{3}=\left[\begin{array}{ccc}
3E_{0}\partial_{\xi} &  & -\frac{E_{0}(1-12E_{0})}{4}\partial_{X\xi}\\
\\
\frac{1-4E_{0}}{16E_{0}}\partial_{X\xi} &  & 3E_{0}\partial_{\xi}
\end{array}\right]
\]
and $\M_{0}$ follows from (\ref{M0}).The source terms $\mathbf{R}_{2}$
and $\mathbf{R}_{3}$ are given in Appendix B. In particular, $\mathbf{R}_{2}$
is a quadratic polynomial of the components of $\mathbf{v}_{1}$,
viz. $E_{1}$ and $\phi_{1}$ and their space derivatives whereas,
$\mathbf{R}_{3}$ is a cubic polynomial of the components of both
$\mathbf{v}_{1}$ and $\mathbf{v}_{2}$ and their space derivatives.
Hereafter, the hierarchy (\ref{mu1}-\ref{mu3}) is solved order by
order by removing the secularities that are condition on the nonlinear
source terms. This is equivalent to imposing the orthogonality of
the right-hand sides of (\ref{mu2}) and (\ref{mu3}) to the the null-space
of the adjoint operator $\M{}_{0}^{*}$.

To $O(\epsilon)$, (\ref{mu1}) is linear and the its general solution
is given by

\[
\mathbf{v}_{1}=\left[\begin{array}{c}
Ae^{i\left(kX-wT\right)}+c.c.\\
\\
\phi_{0}
\end{array}\right],
\]
where the unknown coefficients $A$ and $\phi_{0}$ are function of
the slow scales $\xi$ and $\tau$. Near the linear instability threshold,
$k=k_{c}-q_{e}\epsilon^{2}$ with $q_{e}>0$, and $\mathbf{v}_{1}$
is re-written as
\begin{equation}
\mathbf{v}_{1}=\left[\begin{array}{c}
a(\xi,\tau)e^{i\theta}+c.c.\\
\\
\phi_{0}(\xi,\tau)
\end{array}\right],\label{v1}
\end{equation}
 where $\theta=k_{c}X-w_{c}T$, and the auxiliary amplitude 
\begin{equation}
a(\xi,\tau)=A(\xi,\tau)e^{-iq_{e}\xi}.\label{ax}
\end{equation}
To $O(\epsilon^{2})$, $\mathbf{v}_{2}$ of (\ref{mu2}) is given
by the sum of particular solution $\mathbf{v}_{2,p}$ and the homogenous
solution $\mathbf{v}_{2,h}$ as
\begin{equation}
\mathbf{v}_{2}=\mathbf{v}_{2,p}+\mathbf{v}_{2,h},\label{v2}
\end{equation}
where
\begin{equation}
\mathbf{v}_{2,p}=\left[\begin{array}{c}
q_{0}+q_{12}e^{2i\theta}+c.c.\\
\\
q_{21}e^{i\theta}+q_{22}e^{2i\theta}+c.c.
\end{array}\right],\qquad\mathbf{v}_{2,h}=\left[\begin{array}{c}
\begin{array}{c}
\alpha_{1}e^{i\theta}+c.c.\\
\\
\end{array}\\
\alpha_{3}
\end{array}\right],\label{v2a}
\end{equation}
Here, the unknown coefficients $q_{0}$, $q_{12}$,$q_{21}$, $q_{22}$,
$\alpha_{1}$ and $\alpha_{3}$ are function of the slow scales. Using
MATHEMATICA \cite{MATHEMATICA} coupled with cumbersome algebra shows
that the righthand side of (\ref{mu2}) contains secular terms. Indeed, 

\[
\mathbf{S}_{2}=\left[\begin{array}{c}
\frac{1}{4}e^{i\theta}\left[4\partial_{\tau}a+E_{0}\left(1-12E_{0}\right)k_{c}^{2}q_{21}\right]+c.c.\\
\\
\left(\partial_{\tau}\phi_{0}+\frac{k_{c}^{2}}{16E_{0}^{2}}|a|^{2}a+q_{0}\right)
\end{array}\right]+\mathbf{T,}
\]
and $\mathbf{T}$ contains the non-secular higher-order harmonics
contributions. The secular terms can be removed if $q_{21}$ and $q_{0}$
are chosen such as

\begin{equation}
q_{_{0}}=-\phi_{0\tau}-\frac{k_{c}^{2}}{16E_{0}^{2}}|a|^{2},\qquad q_{21}=-\frac{4a_{\tau}}{E_{0}\left(1-12E_{0}\right)k_{c}^{2}}.\label{q0&q21}
\end{equation}
One can now solve for the non-secular source terms and obtain the
other two coefficients as

\begin{equation}
q_{_{12}}=-\frac{3a^{2}k_{c}}{8E_{0}\left[4E_{0}-k_{c}^{2}\left(1-4E_{c}\right)\right]},\qquad q_{22}=-ia^{2}\varepsilon\frac{12-k_{c}^{2}}{4E_{0}\left(1-12E_{0}\right)k_{c}}.\label{q12&q22}
\end{equation}
The coefficients ($\alpha_{1}$,$\alpha_{3}$) of the homogenous part
together with the amplitude $a$ can be solved at the next order as
follows. To $O(\epsilon^{3})$, substituting the solutions for $v_{1}$
and $v_{2}$ (Eqs. \ref{v1},\ref{v2},\ref{q0&q21},\ref{q12&q22})
into the right-hand side of (\ref{mu3}) yields a source term that
still contains secular terms given by 

\[
\mathbf{S}_{3}=\left[\begin{array}{c}
S_{10}+S_{11}e^{i\theta}+c.c.\\
\\
S_{20}+S_{21}e^{i\theta}+c.c.
\end{array}\right],
\]
where
\begin{equation}
S_{10}=q_{0\tau},\label{S10}
\end{equation}

\begin{equation}
S_{11}=\alpha_{1\tau}+3E_{0}a_{\xi}-3ik_{c}a\phi_{0\tau}+d|a|^{2}a,\label{S11}
\end{equation}

\begin{equation}
S_{20}=\alpha_{3\tau}+3E_{0}\phi_{0\xi}-3ik_{c}\frac{1-4E_{0}}{4E_{0}^{2}\left(1-12E_{0}\right)}\left(a^{*}a_{\tau}-aa_{\tau}^{*}\right),\label{S20}
\end{equation}

\begin{equation}
S_{21}=q_{21\tau}-\frac{k^{2}}{16E_{0}^{2}}a\phi_{0\tau}+ik\frac{1-4\varepsilon^{2}E_{0}}{8E_{0}}a_{\xi}+d_{2}|a|^{2}a,\label{S21}
\end{equation}
and 

\[
d=ik\frac{-48E_{0}\left(1-12E_{0}\right)+k_{c}^{2}\left(-3+108E_{0}+64E_{0}^{2}(-13+24E_{0})\right)}{32E_{0}^{2}\left(1-16E_{0}+48E_{0}^{2}\right)},
\]

\[
d_{2}=-\frac{3-88E_{0}+576E_{0}^{2}-1024E_{0}^{3}}{2E_{0}^{2}\left(1-12E_{0}\right)\left(1-4E_{0}\right)^{2}}.
\]

To have bounded solutions for $\mathbf{v}_{3}$, these secular terms
must be removed. From (\ref{S10}), $S_{10}=0$ yields $q_{0}=c_{0}$,
where the constant $c_{0}$ can be set equal to zero, if at initial
time $q_{0}(\tau=0)=0$. From (\ref{q0&q21}), it then follows that

\[
\phi_{0\tau}=-\frac{k_{c}^{2}}{16E_{0}^{2}}|a|^{2}.
\]
Imposing $S_{11}=0$ and $S_{12}=0$ yield two equations from which
one can solve for $\alpha_{1}$ and $\alpha_{3}$ once $a$ is known.
The evolution equation for $a$ follows from (\ref{S21}) by setting
$S_{21}=0$, which yields

\begin{equation}
i\chi a_{\xi}-a_{\tau\tau}-\beta|a|^{2}a=0,\label{NLSa}
\end{equation}
where $\chi$ is given in (\ref{gamma}) and 

\[
\beta=\frac{2\left(1-8E_{0}\right)\left(1-56E_{0}+128E_{0}^{2}\right)}{\left(1-4E_{0}\right)^{3}}
\]
The main equation (\ref{NLS}) for $A$ follows from (\ref{NLSa})
by substituting $a=Ae^{-iq_{e}\xi}$ {[}see Eq. (\ref{ax}){]} and
the derivative $a_{\xi}=\left(A_{\xi}-iq_{e}\right)e^{-iq_{e}\xi}$
.

\section*{Appendix B: Source terms}

The components of 
\[
\mathbf{R}_{2}=\left[\begin{array}{c}
R_{21}\\
\\
R_{22}
\end{array}\right]
\]
are given by 

\[
\begin{array}{c}
R_{21}=3E_{1}E_{1X}+\frac{1}{4}\left(-1+24E_{0}\right)\left(E_{1}\phi_{1XX}+E_{1X}\phi_{1X}\right)+\\
\\
+\frac{1}{4}E_{1X}\left(E_{1XX}+E_{0}^{2}\phi_{1X}\phi_{1XX}\right),
\end{array}
\]
and 

\[
\begin{array}{c}
R_{22}=\frac{1}{32E_{0}^{2}}\left[-E_{1X}^{2}+E_{1}\left(-2E_{1XX}+96E_{0}^{2}\phi_{1X}\right)+\right.\\
\\
\left.4E_{0}^{2}\left(-1+24E_{0}\right)\phi_{1X}^{2}-8E_{0}^{2}\left(\phi_{1X}^{2}E_{1XX}+E_{1X}\phi_{1XX}\right)\right],
\end{array}
\]
and those of 
\[
\mathbf{R}_{3}=\left[\begin{array}{c}
R_{31}\\
\\
R_{32}
\end{array}\right]
\]
by 

\[
\begin{array}{c}
R_{31}=\frac{1}{4}\left(E_{2X}E_{1XX}+E_{1X}E_{2XX}\right)+\frac{1}{4}\left(-1+24E_{0}\right)\left(\phi_{1X}E_{2X}+E_{2}\phi_{1XX}+E_{1}\phi_{2XX}\right)+\\
\\
3\left(E_{1}^{2}\phi_{1XX}+2E_{1}E_{1X}\phi_{1X}+2E_{0}E_{1X}\phi_{2X}\right)+\\
\\
3\left(E_{0}^{2}\phi_{2X}\phi_{1XX}+E_{0}^{2}\phi_{1X}\phi_{2XX}+E_{1}E_{0}\phi_{1X}\phi_{1XX}+E_{0}E_{1X}\phi_{1X}^{2}\right)+\\
\\
+3\left(E_{1}E_{2X}+E_{2}E_{1X}\right)-\frac{1}{4}E_{1X}\phi_{2X},
\end{array}
\]

and

\[
\begin{array}{c}
R_{33}=\frac{1}{16E_{0}^{3}}\left[E_{1}E_{1X}^{2}+E_{1}^{2}E_{1XX}-E_{0}\left(E_{2}E_{1XX}+E_{1X}E_{2X}+E_{1}E_{2XX}\right)\right]-\frac{\phi_{1X}\phi_{2X}}{4}+\\
\\
3\left(E_{2}\phi_{1X}+E_{1}\phi_{2X}\right)+3\left(E_{1}\phi_{1X}^{2}+2E_{0}\phi_{1X}\phi_{2X}\right)+\\
\\
+\frac{1}{4}\left(E_{0}\phi_{1X}^{3}-E_{2XX}\phi_{1X}-E_{2X}\phi_{1XX}-E_{1XX}\phi_{2X}-E_{1X}\phi_{2XX}\right).
\end{array}
\]

\section{Appendix C: Multiple-scale perturbation analysis for $\mu>\mu_{c}$}

We draw on \cite{TaniutiWeiKdV} and define the multiple scales $\xi=\epsilon(X-cT),\:\tau=\epsilon^{2}T$,
with c as a wave celerity to be determined. Substitution of the ordered
expansion (\ref{w}) into the set of the main equations (\ref{K_Eeq})
for $E$ and $K$ yields the following hierarchy of vector equations 

\begin{equation}
O(\epsilon^{2}):\qquad(-c\mathbf{I}+\mathbf{V})\mathbf{\partial_{\xi}w}_{1}=0\label{wmu1}
\end{equation}

\begin{equation}
O(\epsilon^{3}):\qquad(-c\mathbf{I}+\mathbf{V})\mathbf{\partial_{\xi}w}_{2}=-\partial_{\tau}\mathbf{w}_{1}+\mathbf{R}(\mathbf{w}_{1}),\label{wmu2}
\end{equation}
where 
\begin{equation}
\mathbf{V}=\left[\begin{array}{ccc}
3E_{0} &  & -\frac{E_{0}(1-12E_{0})}{4}\\
\\
1 &  & 3E_{0}
\end{array}\right]\label{vm}
\end{equation}
and the source term
\[
\mathbf{R}=-\left[\begin{array}{c}
3K_{1}K_{1\xi}+\frac{-1+24E_{0}}{4}K_{1}E_{1\xi}\\
\\
3(KE_{1\xi}+E_{1}K_{1\xi})+\frac{-1+24E_{0}}{4}K_{1}K_{1\xi}
\end{array}\right]
\]
The eigenvalues of $\mathbf{V}$ follows as 
\[
\lambda_{1,2}=3E_{0}\pm\frac{1}{2}\sqrt{-1+12E_{0}}.
\]
and the associated right- and left- eigenvectors are given, respectively,
by 
\[
\mathbf{q}_{1,2}=\left[\begin{array}{c}
\pm\frac{\sqrt{-1+12E_{0}}}{2}\\
\\
1
\end{array}\right],\qquad\mathbf{p}_{1,2}=\left[\begin{array}{ccc}
\pm\frac{2}{\sqrt{-1+12E_{0}}} &  & 1\end{array}\right].
\]
 The eigenvalues are real for $E_{0}\geq E_{c}$ denoting the hyperbolic
nature of hierarchy equations; however $\mathbf{V}$ is diagonalizable
only for $E_{0}>E_{c}$ (For $E_{0}=E_{c}$, $\mathbf{V}$ can be
made triangular via the Jordan decomposition, but this case will not
be considered here). As a result, to $O(\epsilon^{2})$ we set $c=\lambda_{j}$
and the solution of (\ref{wmu1}) is given by 
\begin{equation}
\mathbf{w}_{1}=\mathbf{q}_{j}F(\xi,\tau),\label{ww1}
\end{equation}
and $F$ is solved to the next order. Indeed, to $O(\epsilon^{3})$
the compatibility condition for (\ref{wmu2}) imposes its source term
to be orthogonal to the corresponding row left-eigenvector $\mathbf{p}_{j}$
of $\mathbf{V}$ (see \cite{TaniutiWeiKdV}). This yields 
\[
F_{\tau}-\frac{\mathbf{p}_{j}R(\mathbf{q}_{j}F)}{\mathbf{p}_{j}\mathbf{q}_{j}}=0,
\]
which after some simplifications yields the non-dispersive KdV equation
(\ref{KdV}).

\section{Appendix D: Excess kurtosis}

Drawing upon \cite{Janssen2006}, the excess kurtosis of weakly nonlinear
waves that obey the local cDZ equation is given by

\begin{equation}
\lambda_{40}(t)=\frac{24\mu^{2}}{\nu^{2}}\iiint T_{123}\sqrt{\frac{w}{w_{1}w_{2}w_{3}}}S_{1}S_{2}S_{3}\frac{1-\cos(\Delta\omega_{0}t)}{\Delta}dk_{1}dk_{2}dk_{3,}\label{mu40}
\end{equation}
where the dimensionless frequency $w_{j}=\sqrt{k_{j}}$, $\Delta=w+w_{1}-w_{2}-w_{3}$,
the dimensionless Gaussian spectra 
\[
S_{j}(k_{j})=\frac{1}{\sqrt{2\pi}}e^{-\frac{(k_{j}-1)^{2}}{2\nu^{2}}},
\]
and the kernel $T_{123}$ is given in \cite{Dyachenko2011} as 
\[
T_{123}=\frac{1}{8\pi}\left[kk_{1}\left(k+k_{1}\right)+k_{2}k_{3}\left(k_{2}+k_{3}\right)\right].
\]
To solve for (\ref{mu40}), define the auxiliary variables $z_{j}=(k_{j}-1)/\nu$.
Then, correct to $O(\nu^{2})$, Eq. (\ref{mu40}) reduces to 

\begin{equation}
\lambda_{40}=\frac{24\mu^{2}}{\nu^{2}}\iiint\frac{e^{-\frac{z_{1}^{2}+z_{2}^{2}+z_{3}^{2}}{2\nu^{2}}}}{(2\pi)^{3/2}}G\frac{1-\cos(Z\alpha)}{Z/4}dz_{1}dz_{2}dz_{3,}\label{mu4}
\end{equation}
where $\alpha=\frac{1}{4}\nu^{2}\omega_{0}t$, $Z=-\left(z_{1}-z_{2}\right)\left(z_{1}-z_{3}\right)$,
and

\[
G=1+\frac{\nu}{2}\left(-z_{1}+3z_{2}+3z_{3}\right)+\frac{\nu^{2}}{8}\left(-3z_{1}^{2}+4z_{2}^{2}+10z_{2}z_{3}+4z_{3}^{2}\right).
\]
Following \cite{fedeleNLS}, 

\[
\frac{d\lambda_{40}}{d\alpha}=\frac{24\mu^{2}}{\nu^{2}}\iiint\frac{e^{-\frac{z_{1}^{2}+z_{2}^{2}+z_{3}^{2}}{2\nu^{2}}}}{(2\pi)^{3/2}}G\sin(Z\alpha)dz_{1}dz_{2}dz_{3,}
\]
and in vector notation

\begin{equation}
\frac{d\lambda_{40}}{d\alpha}=\frac{24\mu^{2}}{\nu^{2}}\textrm{\ensuremath{\Im}}\left[J(\alpha)\right],\label{dmu}
\end{equation}
where $\Im\left(x\right)$ denotes the imaginary part of $x$,

\begin{equation}
J(\alpha)=\iiint\frac{e^{-\frac{1}{2}\mathbf{z}^{T}\mathbf{\Omega}\mathbf{z}}}{(2\pi)^{3/2}}\left(1+\frac{\nu}{2}\mathbf{c}^{T}\mathbf{z}+\frac{\nu^{2}}{8}\mathbf{z}^{T}\mathbf{A}\mathbf{z}\right)d\mathbf{z,}\label{J}
\end{equation}
and 
\[
\mathbf{z}=\left[\begin{array}{c}
z_{1}\\
z_{2}\\
z_{3}
\end{array}\right],\quad\mathbf{c}=\left[\begin{array}{c}
-1\\
3\\
3
\end{array}\right],
\]
\[
\mathbf{\Omega}=\left[\begin{array}{ccc}
1+2i\alpha & -i\alpha & -i\alpha\\
-i\alpha & 1 & i\alpha\\
-i\alpha & i\alpha & 1
\end{array}\right],\qquad\mathbf{A}=\left[\begin{array}{ccc}
-3 & 0 & 0\\
0 & 4 & 5\\
0 & 5 & 4
\end{array}\right],
\]
The Gaussian integral (\ref{J}) can be solved exactly by the change
of variable $\mathbf{s}=\mathbf{Q}^{-1}\mathbf{z}$, where $\mathbf{Q}$
is the eigenvector matrix of $\mathbf{\Omega}=\mathbf{Q}^{-1}\mathbf{D}\mathbf{Q}$,
and $\mathbf{D}$ that of the eigenvalues. Then,

\[
J(\alpha)=\iiint\frac{e^{-\frac{1}{2}\mathbf{s}^{T}\mathbf{D}\mathbf{s}}}{(2\pi)^{3/2}}\left(1+\frac{\nu^{2}}{8}\mathbf{s}^{T}\mathbf{\mathbf{Q}^{-T}\mathbf{A}\mathbf{Q}^{-1}}\mathbf{s}\right)d\mathbf{s,}
\]
and after integration 

\[
J(\alpha)=\frac{24+48i\alpha+72\alpha^{2}+\nu^{2}\left(5+10i\alpha-9\alpha^{2}\right)}{24\left(1+2i\alpha+3\alpha^{2}\right)^{3/2}}.
\]
Integrating (\ref{dmu}) in $\alpha$ with $J(0)=0$ yields

\[
\lambda_{40}^{cDZ}=\frac{24\mu^{2}}{\nu^{2}}\left[\left(\frac{\pi}{6\sqrt{3}}-\nu^{2}\frac{12+\pi\sqrt{3}}{144}\right)-\Im\left(\frac{2\nu^{2}\left(5\alpha-i\right)+i\sqrt{3}\left(8-\nu^{2}\right)\sqrt{1+2i\alpha+9\alpha^{2}}\arcsin\left(\frac{1-3i\alpha}{2}\right)}{\sqrt{1+2i\alpha+3\alpha^{2}}}\right)\right].
\]
At steady state, as $\alpha\rightarrow\infty,$

\[
\lambda_{40}^{cDZ}=\frac{24\mu^{2}}{\nu^{2}}\left(\frac{\pi}{6\sqrt{3}}-\nu^{2}\frac{12+\pi\sqrt{3}}{144}\right).
\]
The NLS limit derived by \cite{Janssen2006} follows by neglecting
$O(\nu^{2})$ terms within parenthesis, namely

\[
\lambda_{40}^{NLS}=\frac{\pi}{6\sqrt{3}}\frac{24\mu^{2}}{\nu^{2}}.
\]
Thus, correct to $O(\nu^{2})$, 
\[
\lambda_{40}^{cDZ}=\lambda_{40}^{NLS}\left(1-\frac{4\sqrt{3}+\pi}{8\pi}\nu^{2}\right)\approx\lambda_{40}^{NLS}\left(1-0.40\nu^{2}\right).
\]

\bibliographystyle{jfm}
\bibliography{biblioFranco}

\begin{thebibliography}{53}
\expandafter\ifx\csname natexlab\endcsname\relax\def\natexlab#1{#1}\fi

\bibitem[8.0(2010)]{MATHEMATICA}
{\sc 8.0, Mathematica} 2010 Wolfram research, inc. .

\bibitem[Ablowitz \& Segur(1981)]{Ablowitz1981}
{\sc Ablowitz, M.~J. \& Segur, H.} 1981 {\em {Solitons and the Inverse
  Scattering Transform}\/}. Society for Industrial \& Applied Mathematics.

\bibitem[Baldock {\em et~al.\/}(1996)Baldock, Swan \& Taylor]{Taylor1996}
{\sc Baldock, T.~E., Swan, C. \& Taylor, P.~H.} 1996 A laboratory study of
  nonlinear surface waves on water. {\em Philosophical Transactions of the
  Royal Society of London. Series A: Mathematical, Physical and Engineering
  Sciences\/} {\bf 354}~(1707), 649--676.

\bibitem[Banner {\em et~al.\/}(2013)Banner, Barthelemy, Fedele, Allis,
  Benetazzo, Dias \& Peirson]{Banneretal2013}
{\sc Banner, M.~L., Barthelemy, X., Fedele, F., Allis, M., Benetazzo, A., Dias,
  F. \& Peirson, W.L.} 2013 Unexpected wave group behaviour challenges use of
  stokes theory for ocean waves ~(http://arxiv.org/abs/1305.3980).

\bibitem[Benjamin(1967)]{Benjamin1967}
{\sc Benjamin, T.~B.} 1967 {Instability of periodic wavetrains in nonlinear
  dispersive systems}. {\em Proc. R. Soc. London, Ser. A\/} {\bf 299(59)}.

\bibitem[Benjamin \& Feir(1967)]{Benjamin1967a}
{\sc Benjamin, T.~B. \& Feir, J.~E.} 1967 {The disintegration of wavetrains in
  deep water. Part 1}. {\em J. Fluid Mech.\/} {\bf 27}~(417).

\bibitem[Bridges(2004)]{BridgesJFMhomoclinic}
{\sc Bridges, Thomas~J.} 2004 Superharmonic instability, homoclinic torus
  bifurcation and water-wave breaking. {\em Journal of Fluid Mechanics\/} {\bf
  505}, 153--162.

\bibitem[Bridges(2013)]{BridgesKdV2013}
{\sc Bridges, Thomas~J.} 2013 A universal form for the emergence of the
  korteweg--de vries equation. {\em Proceedings of the Royal Society A:
  Mathematical, Physical and Engineering Science\/} {\bf 469}~(2153).

\bibitem[Camassa \& Holm(1993)]{Camassa1993}
{\sc Camassa, R. \& Holm, D.} 1993 {An integrable shallow water equation with
  peaked solitons}. {\em Phys. Rev. Lett.\/} {\bf 71(11)}, 1661--1664.

\bibitem[Chabchoub {\em et~al.\/}(2012)Chabchoub, Hoffmann, Onorato \&
  Akhmediev]{Chabchoubc2012}
{\sc Chabchoub, A., Hoffmann, N., Onorato, M. \& Akhmediev, N.} 2012 Super
  rogue waves: Observation of a higher-order breather in water waves. {\em
  Phys. Rev. X\/} {\bf 2}, 011015.

\bibitem[Chabchoub {\em et~al.\/}(2011)Chabchoub, Hoffmann \&
  Akhmediev]{Chabchoub2011}
{\sc Chabchoub, A., Hoffmann, N.~P. \& Akhmediev, N.} 2011 Rogue wave
  observation in a water wave tank. {\em Phys. Rev. Lett.\/} {\bf 106}, 204502.

\bibitem[Chu \& Mei(1970)]{Chu1970}
{\sc Chu, V.~H. \& Mei, C.~C.} 1970 {On slowly-varying Stokes waves}. {\em
  Journal of Fluid Mechanics\/} {\bf 41}, 873--887.

\bibitem[Clamond {\em et~al.\/}(2006)Clamond, Francius, Grue \&
  Kharif]{Clamond2006}
{\sc Clamond, D., Francius, M., Grue, J. \& Kharif, C.} 2006 {Long time
  interaction of envelope solitons and freak wave formations}. {\em Eur. J.
  Mech. B/Fluids\/} {\bf 25}~(5), 536--553.

\bibitem[Crawford {\em et~al.\/}(1981)Crawford, Lake, Saffman \&
  Yuen]{Crawfordetal1981}
{\sc Crawford, Donald~R., Lake, Bruce~M., Saffman, Philip~G. \& Yuen, Henry~C.}
  1981 Stability of weakly nonlinear deep-water waves in two and three
  dimensions. {\em Journal of Fluid Mechanics\/} {\bf 105}, 177--191.

\bibitem[Dyachenko \& Zakharov(2011)]{Dyachenko2011}
{\sc Dyachenko, A.~I. \& Zakharov, V.~E.} 2011 {Compact Equation for Gravity
  Waves on Deep Water}. {\em JETP Lett.\/} {\bf 93}~(12), 701--705.

\bibitem[Dyachenko \& Zakharov(2013)]{Dyachenko2013}
{\sc Dyachenko, A.~I., Kachulin D.~I. \& Zakharov, V.~E.} 2013 On the
  nonintegrability of the free surface hydrodynamics. {\em JETP Lett.\/} {\bf
  98}~(1), 48--52.

\bibitem[Dysthe(1979)]{Dysthe1979}
{\sc Dysthe, K.~B.} 1979 {Note on a modification to the nonlinear
  Schr\"{o}dinger equation for application to deep water}. {\em Proc. R. Soc.
  Lond. A\/} {\bf 369}, 105--114.

\bibitem[Dysthe \& Trulsen(1999)]{Dysthe1999}
{\sc Dysthe, K.~B. \& Trulsen, K.} 1999 {Note on breather type solutions of the
  NLS as models for freak-waves}. {\em Physica Scripta\/} {\bf T82}, 48--52.

\bibitem[Dysthe \& Muller(2008)]{DystheKrogstad2008}
{\sc Dysthe, K.~B., Krogstad H.~E. \& Muller, P.} 2008 Oceanic rogue waves.
  {\em Annual Review of Fluid Mechanics\/} {\bf 40}, 287--310.

\bibitem[Fedele {\em et~al.\/}(2010)Fedele, Cherneva, Tayfun \&
  Soares]{fedeleNLS}
{\sc Fedele, F., Cherneva, Z., Tayfun, M.~A. \& Soares, C.~Guedes} 2010
  Nonlinear schrodinger invariants and wave statistics. {\em Physics of
  Fluids\/} {\bf 22}~(3), 036601.

\bibitem[Fedele \& Dutykh(2012)]{FedeleDutykhJFM2012}
{\sc Fedele, F. \& Dutykh, D.} 2012 {Special solutions to a compact equation
  for deep-water gravity waves}. {\em Journal of Fluid Mechanics\/} {\bf 712},
  646--660.

\bibitem[Fermi(1955)]{FermiPastaUlam}
{\sc Fermi, E., Pasta J. Ulam~H.C.} 1955 Studies of non linear problems. {\em
  Tech. Rep.\/} Report No. LA-1940. Los Alamos Scientific Laboratory.

\bibitem[Gramstad \& Trulsen(2007)]{GramstadTrulsen2007}
{\sc Gramstad, O. \& Trulsen, K.} 2007 Influence of crest and group length on
  the occurrence of freak waves. {\em Journal of Fluid Mechanics\/} {\bf 582},
  463--472.

\bibitem[Henderson {\em et~al.\/}(1999)Henderson, Peregrine \&
  Dold]{Henderson1999341}
{\sc Henderson, K.L., Peregrine, D.H. \& Dold, J.W.} 1999 Unsteady water wave
  modulations: fully nonlinear solutions and comparison with the nonlinear
  schr{\"o}dinger equation. {\em Wave Motion\/} {\bf 29}~(4), 341 -- 361.

\bibitem[Janssen(1981)]{JanssenPoF1981}
{\sc Janssen, P.A.E.M.} 1981 Modulational instability and the fermi-pasta-ulam
  recurrence. {\em Physics of Fluids\/} {\bf 24}, 23--26.

\bibitem[Janssen(1983)]{Janssen1983}
{\sc Janssen, P.A.E.M.} 1983 {On a fourth-order envelope equation for
  deep-water waves}. {\em J. Fluid Mech\/} {\bf 126}, 1--11.

\bibitem[Janssen(2003)]{Janssen2003}
{\sc Janssen, Peter A. E.~M.} 2003 Nonlinear four-wave interactions and freak
  waves. {\em Journal of Physical Oceanography\/} {\bf 33}~(4), 863--884.

\bibitem[Jillians(1989)]{Jillians}
{\sc Jillians, W.~J.} 1989 The superharmonic instability of stokes waves in
  deep water. {\em Journal of Fluid Mechanics\/} {\bf 204}, 563--579.

\bibitem[Kharif \& Pelinovsky(2003)]{Kharif2003}
{\sc Kharif, C. \& Pelinovsky, E.} 2003 {Physical mechanisms of the rogue wave
  phenomenon}. {\em Eur. J. Mech. B/Fluids\/} {\bf 22}, 603--634.

\bibitem[Kharif {\em et~al.\/}(2009)Kharif, Pelinovsky \&
  Slunyaev]{Kharif2009a}
{\sc Kharif, C., Pelinovsky, E. \& Slunyaev, A.} 2009 {\em {Rogue Waves in the
  Ocean}\/}. Springer.

\bibitem[Krasitskii(1994)]{Krasitskii1994}
{\sc Krasitskii, V.~P.} 1994 {On reduced equations in the Hamiltonian theory of
  weakly nonlinear surface waves}. {\em J. Fluid Mech\/} {\bf 272}, 1--20.

\bibitem[Lake {\em et~al.\/}(1977)Lake, Yuen, Rungaldier \& Ferguson]{lake}
{\sc Lake, Bruce~M., Yuen, Henry~C., Rungaldier, Harald \& Ferguson, Warren~E.}
  1977 Nonlinear deep-water waves: theory and experiment. part 2. evolution of
  a continuous wave train. {\em Journal of Fluid Mechanics\/} {\bf 83}, 49--74.

\bibitem[Lighthill(1965)]{LIGHTHILL01091965}
{\sc Lighthill, M.~J.} 1965 Contributions to the theory of waves in non-linear
  dispersive systems. {\em IMA Journal of Applied Mathematics\/} {\bf 1}~(3),
  269--306.

\bibitem[Longuet-Higgins(1978{\natexlab{{\em a\/}}})]{Longuet-HigginspartI1978}
{\sc Longuet-Higgins, M.~S.} 1978{\natexlab{{\em a\/}}} The instabilities of
  gravity waves of finite amplitude in deep water. i. superharmonics. {\em
  Proceedings of the Royal Society of London. A. Mathematical and Physical
  Sciences\/} {\bf 360}~(1703), 471--488.

\bibitem[Longuet-Higgins(1978{\natexlab{{\em
  b\/}}})]{Longuet-HigginspartII1978}
{\sc Longuet-Higgins, M.~S.} 1978{\natexlab{{\em b\/}}} The instabilities of
  gravity waves of finite amplitude in deep water ii. subharmonics. {\em
  Proceedings of the Royal Society of London. A. Mathematical and Physical
  Sciences\/} {\bf 360}~(1703), 489--505.

\bibitem[Longuet-Higgins \& Cokelet(1978)]{Longuet-HigginsCokelet1978}
{\sc Longuet-Higgins, M.~S. \& Cokelet, E.~D.} 1978 The deformation of steep
  surface waves on water. ii. growth of normal-mode instabilities. {\em
  Proceedings of the Royal Society of London. A. Mathematical and Physical
  Sciences\/} {\bf 364}~(1716), 1--28.

\bibitem[McLean(1982)]{McLean1982}
{\sc McLean, John~W.} 1982 Instabilities of finite-amplitude water waves. {\em
  Journal of Fluid Mechanics\/} {\bf 114}, 315--330.

\bibitem[Mori \& Janssen(2006)]{Janssen2006}
{\sc Mori, Nobuhito \& Janssen, Peter A. E.~M.} 2006 On kurtosis and occurrence
  probability of freak waves. {\em Journal of Physical Oceanography\/} {\bf
  36}~(7), 1471--1483.

\bibitem[Osborne(2010)]{Osborne2010}
{\sc Osborne, A.} 2010 {\em {Nonlinear ocean waves and the inverse scattering
  transform}\/}, , vol.~97. Elsevier.

\bibitem[Osborne {\em et~al.\/}(2000)Osborne, Onorato \& Serio]{Osborne2000}
{\sc Osborne, A.~R., Onorato, M. \& Serio, M.} 2000 {The nonlinear dynamics of
  rogue waves and holes in deep-water gravity wave trains}. {\em Phys. Lett.
  A\/} {\bf 275}~(5-6), 386--393.

\bibitem[Peregrine(1983)]{Peregrine1983}
{\sc Peregrine, D.~H.} 1983 {Water waves, nonlinear Schr\"{o}dinger equations
  and their solutions}. {\em Journal of the Australian Mathematical Society
  Series B\/} {\bf 25}, 16--43.

\bibitem[Rapp \& Melville(1990)]{RappMelville}
{\sc Rapp, R.~J. \& Melville, W.~K.} 1990 Laboratory measurements of deep-water
  breaking waves. {\em Philosophical Transactions of the Royal Society of
  London. Series A, Mathematical and Physical Sciences\/} {\bf 331}~(1622),
  735--800.

\bibitem[Schober(2006)]{Schober2006}
{\sc Schober, C.M.} 2006 Melnikov analysis and inverse spectral analysis of
  rogue waves in deep water. {\em European Journal of Mechanics - B/Fluids\/}
  {\bf 25}~(5), 602 -- 620.

\bibitem[Shemer \& Alperovich(2013)]{ShemerPoF2013}
{\sc Shemer, L. \& Alperovich, S.H.} 2013 Peregrine breather revisited. {\em
  Physics of Fluids\/} {\bf 25}, 051701.

\bibitem[Slunyaev {\em et~al.\/}(2013)Slunyaev, Pelinovsky, Sergeeva,
  Chabchoub, Hoffmann, Onorato \& Akhmediev]{Slunyaev2013PRE}
{\sc Slunyaev, A., Pelinovsky, E., Sergeeva, A., Chabchoub, A., Hoffmann, N.,
  Onorato, M. \& Akhmediev, N.} 2013 Super-rogue waves in simulations based on
  weakly nonlinear and fully nonlinear hydrodynamic equations. {\em Phys. Rev.
  E\/} {\bf 88}, 012909.

\bibitem[Slunyaev \& Shrira(2013)]{SlunyaevShrira2013}
{\sc Slunyaev, Alexey~V. \& Shrira, Victor~I.} 2013 On the highest non-breaking
  wave in a group: fully nonlinear water wave breathers versus weakly nonlinear
  theory. {\em Journal of Fluid Mechanics\/} {\bf 735}, 203--248.

\bibitem[Sulem \& Sulem(1999)]{Sulem1999}
{\sc Sulem, C. \& Sulem, P.-L.} 1999 {\em {The Nonlinear Schr\"{o}dinger
  Equation. Self-Focusing and Wave Collapse}\/}. Springer-Verlag, New York.

\bibitem[Tanaka(1990)]{Tanaka1990559}
{\sc Tanaka, Mitsuhiro} 1990 Maximum amplitude of modulated wavetrain. {\em
  Wave Motion\/} {\bf 12}~(6), 559 -- 568.

\bibitem[Tanaka {\em et~al.\/}(1987)Tanaka, Dold, Lewy \& Peregrine]{Tanaka}
{\sc Tanaka, M., Dold, J.~W., Lewy, M. \& Peregrine, D.~H.} 1987 Instability
  and breaking of a solitary wave. {\em Journal of Fluid Mechanics\/} {\bf
  185}, 235--248.

\bibitem[Taniuti \& Wei(1968)]{TaniutiWeiKdV}
{\sc Taniuti, Tosiya \& Wei, Chau-Chin} 1968 Reductive perturbation method in
  nonlinear wave propagation. i. {\em Journal of the Physical Society of
  Japan\/} {\bf 24}~(4), 941--946.

\bibitem[Yang(2010)]{Yang2010}
{\sc Yang, J.} 2010 {\em {Nonlinear Waves in Integrable and Nonintegrable
  Systems}\/}. SIAM.

\bibitem[Zakharov(1999)]{Zakharov1999}
{\sc Zakharov, V.~E.} 1999 {Statistical theory of gravity and capillary waves
  on the surface of a finite-depth fluid}. {\em Eur. J. Mech. B/Fluids\/} {\bf
  18}~(3), 327--344.

\bibitem[Zakharov \& Shabat(1972)]{Zakharov1972}
{\sc Zakharov, V.~E. \& Shabat, A.~B.} 1972 {Exact Theory of Two-dimensional
  Self-focusing and One-dimensional Self-modulation of Waves in Nonlinear
  Media}. {\em Soviet Physics-JETP\/} {\bf 34}, 62--69.

\end{thebibliography}

\end{document}